\documentclass{PoS}

\usepackage{graphicx}
\usepackage{amsmath} 
\title{CP violation in top-quark pair production and decay.}

\ShortTitle{CP violation in top-quark pair production and decay.}

\author{\speaker{German Valencia}%
         \thanks{ Work supported in part by DOE under contract number DE-FG02-01ER41155.}\\
        Iowa State University\\
        E-mail: \email{valencia@iastate.edu}}

\abstract{As the LHC becomes a top-quark factory and enters the era of precision measurements it presents us with a unique opportunity to search for new sources of CP violation. I discuss T-odd triple product correlations and the special role they play in the search for CP violation in top physics. I motivate the searches by discussing two examples of new physics that induce these CP violating observables at the LHC.

}

\FullConference{The XIth International Conference on Heavy Quarks and Leptons \\
                 11-15 June, 2012\\
                 Prague,The Czech Republic}

\begin{document}

\section{Introduction}

The LHC is quickly becoming a top-quark factory permitting detailed tests of top-quark interactions. An important aspect of this program should be the search for new sources of CP violation. CP violation beyond the standard model (SM) has yet to be observed but we suspect it must be there in order to explain the baryon asymmetry of the universe. 

Top-quark pair production and decay provides a unique opportunity to study triple product momentum correlations since this process  generates a large chain of linearly independent four-momentum vectors that are correlated by spin \cite{tprods}. The new observables that can be studied are simple kinematic correlations of the form $\vec{p}_1\cdot(\vec{p}_2\times\vec{p}_3)$. 

These correlations are referred to as ``naive-T'' odd because they  reverse sign under the ``naive-T'' operation that reverses the direction of momenta and spin without interchanging initial and final states. These correlations do not have to be CP-odd, they can be induced by CP conserving interactions because the naive-T operation does not correspond to the time reversal operation. It is well known, however, that CP conserving T-odd correlations only occur {\it beyond tree level}, and for this reason we will refer to them as being induced by ``unitarity phases''. This means that the CP conserving background is both small and interesting in its own right. The CP nature of a given correlation can be determined easily as we will show with examples later on. An important point for collider physics is that the generic momenta $\vec{p}_i$ that enters the correlation can be that of a composite object, such as a jet. 

\begin{figure}
\hspace{0.7in}\includegraphics[width=4in]{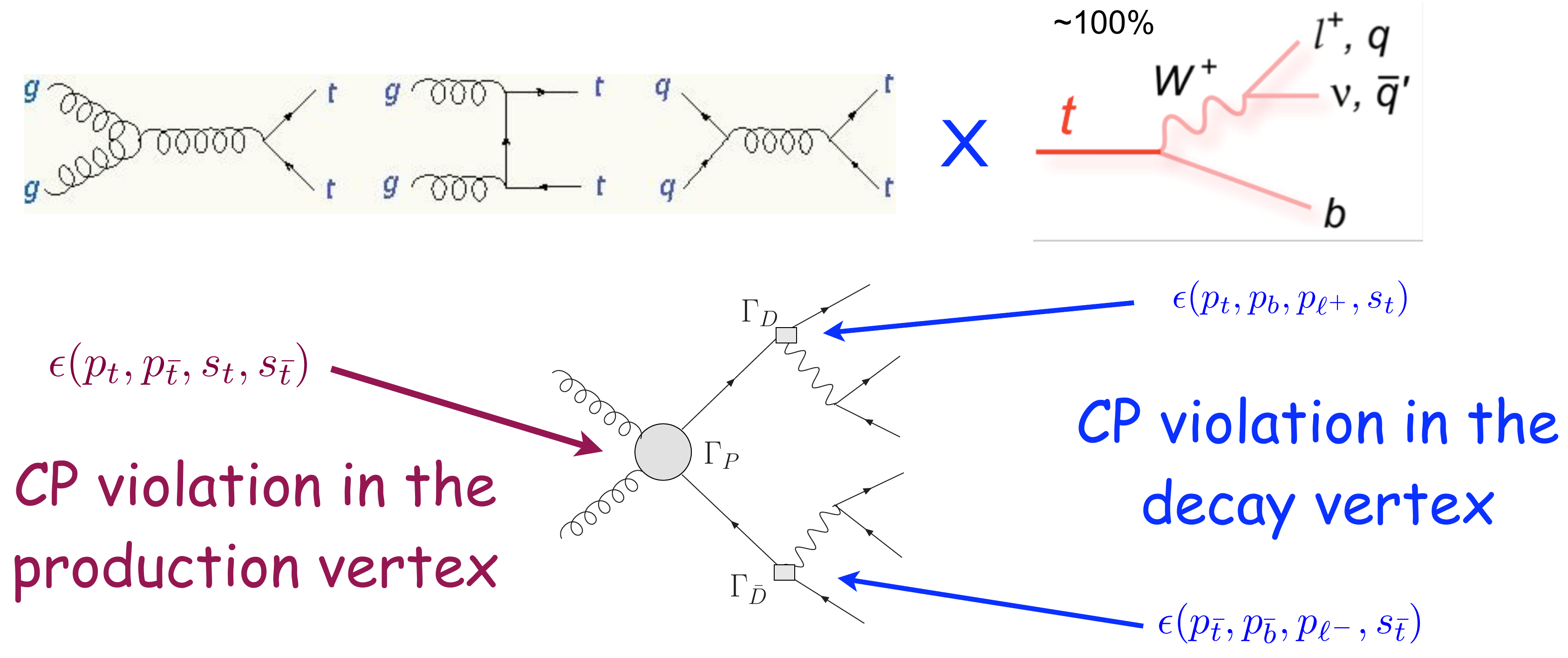}   
\caption{Kinematic configuration: the top-quark and $W$-bosons are treated on-shell.  The T-odd correlations originate as spin correlations due to CP violating new physics in the production and/or decay vertices}
\label{fig1}
\end{figure}

A generic diagram for the processes we consider is shown in Figure~\ref{fig1}. The circle in the production process represents the top-quark pair production including the SM diagrams and CP violation due to new physics. Similarly the square in the top decay process represents top decay via the SM and any additional CP violating interactions. The $W$ decay is assumed to proceed as in the SM and we will consider  both the leptonic and hadronic (jets) cases.
In Fig.~\ref{fig1} we see how the topology of these processes  implies that the T-odd correlations first appear as spin correlations. When CP violation occurs in the production vertex, the only Lorentz invariant (scalar) T-odd correlation that occurs is one that involves the momenta and spin of both the top and anti-top. The weak decay of the top-quark  acts as a spin analyzer and the top-spin gets replaced by one of the final state momenta in the final correlation. In the case of CP violation in the decay vertex we see that the induced spin correlations do not have definite CP properties. A comparison of both decay vertices is required to separate CP violation from signals induced by unitarity phases.

\section{CP-odd correlations}

We first consider a simple example in the dimuon channel as illustration:  $pp \to t\bar{t} \to b\bar{b} \mu^+ \nu \mu^- \bar\nu$. A T-odd correlation that is also CP odd for this case is
\begin{eqnarray}
{\cal O}_1\ =\ \epsilon(p_b,p_{\bar {b}},p_{\mu^+},p_{\mu^-}) \,\,
\xrightarrow[]{(b\bar{b})_{C.M.}}\,\, \propto  \vec{p}_b\
\cdot  \left(\vec{p}_{\mu^+}\times\vec{p}_{\mu^-}\right) \nonumber \\
\xrightarrow[]{CP}\,\, -\vec{p}_{\bar b}\
\cdot  \left(-\vec{p}_{\mu^-}\times-\vec{p}_{\mu^+}\right) \,\, = \,\,
-\vec{p}_b\
\cdot  \left(\vec{p}_{\mu^+}\times\vec{p}_{\mu^-}\right).
\label{samplecor}
\end{eqnarray}
In Eq.~\ref{samplecor} we first write the Lorentz scalar form of the triple product which can be evaluated in any reference frame. To understand its CP properties we next go to the $(b\bar{b})$ center of mass frame as indicated by the arrow in the first line. This results in the familiar triple-product form. The second line sketches the proof that this correlation is indeed CP odd. Notice that this happens even though the LHC is a pp collider because we have treated it as a gluon collider (or $q\bar{q}$ collider) as is appropriate for top-quark pair production. This picture fails only for $qq$ initiated processes which have been estimated to contribute a negligible background to $t\bar{t}$ production \cite{review}. Once we have a correlation such as Eq.~\ref{samplecor}, we can look for CP violation in two ways: by finding asymmetries in the distributions $d\sigma/d{\cal O}_1$, or by constructing an integrated counting asymmetry of the form:
\begin{eqnarray}
A_{CP}=\frac{N_{events}(\vec{p}_b\
\cdot  \left(\vec{p}_{\mu^+}\times\vec{p}_{\mu^-}\right)>0)-N_{events}(\vec{p}_b\
\cdot  \left(\vec{p}_{\mu^+}\times\vec{p}_{\mu^-}\right)<0)}{N_{events}(\vec{p}_b\
\cdot  \left(\vec{p}_{\mu^+}\times\vec{p}_{\mu^-}\right)>0)+N_{events}(\vec{p}_b\
\cdot  \left(\vec{p}_{\mu^+}\times\vec{p}_{\mu^-}\right)<0)}.
\label{counting}
\end{eqnarray}
We have constructed the counting asymmetry Eq.~\ref{counting} in the $b\bar{b}$ center of mass frame, but we could have done  it in any frame, including the lab frame, by using the original Lorentz covariant form of ${\cal O}_1$. At first glance, ${\cal O}_1$ appears to require distinguishing the $b$-jet from the $\bar{b}$-jet. This is not the case, as it suffices to associate each $b$-jet with a given muon using any CP-blind scheme.  A detailed study of a correlation proportional to ${\cal O}_1$ for Atlas has been performed by J. Sj\"{o}lin \cite{Sjolin:2003ah}.

Many correlations can be constructed that are appropriate for different situations. Below we give examples appropriate for dilepton, lepton plus jets, and hadronic top-decay channels; as well as for detecting CP violation or unitarity phases. We will only use the following momenta: lepton ($p_{\mu^\pm}$); $b$-jet ($p_{b,\bar{b}}$); beam momentum ($\tilde{q}\equiv P_1-P_2$); non-$b$ jet momenta ordered by $p_T$ ($p_{j1},p_{j2}\cdots$). Any other CP blind ordering of the non-$b$ jets will also work. 

\begin{itemize}

\item Dimuon events at LHC: CP-odd correlations ${\cal O}_1$ given above and 
\begin{eqnarray}
{\cal{O}}_2 &=& \, \tilde{q}\cdot (p_{\mu^+}-p_{\mu^-}) \,\epsilon(p_{\mu^+},p_{\mu^-},p_b+p_{\bar{b}},\tilde{q}) 
\end{eqnarray}
In this example  it is explicitly not necessary to distinguish the $b$ and $\bar{b}$ jets; it is also quadratic in $\tilde{q}$ as needed for identical particles in the initial state.

\item Dimuon events at LHC: CP-even T-odd correlation to study absorptive phases (so this one does {\it not} signal CP violation):
\begin{eqnarray}
{\cal O}_b &=& \, \tilde{q}\cdot (p_{\mu^+}-p_{\mu^-}) \,\epsilon(p_{\mu^+},p_{\mu^-},p_b-p_{\bar{b}},\tilde{q}).
\end{eqnarray}

\item Muon plus jets events, CP-odd correlations. Notice some require distinguishing $b$ from $\bar{b}$ but some don't.
\begin{eqnarray}
{\cal O}_{2\ell j} &=& \epsilon(P,p_b+p_{\bar b},p_\ell,p_{j1}) \,\,
\xrightarrow[]{lab}\,\, \propto \,\, (\vec{p}_b +\vec{p}_{\bar b})\cdot (\vec{p}_\ell \times \vec{p}_{j1})
\nonumber \\
{\cal O}_{3\ell j} &=&  Q_\ell \, \epsilon(p_b,p_{\bar b},p_\ell,p_{j1})  \,\,
\xrightarrow[]{b\bar b ~CM}\,\, \propto \,\, Q_\ell\,\vec{p}_b \cdot (\vec{p}_\ell \times \vec{p}_{j1})
\nonumber \\
{\cal O}_{7\ell j} &=&  \tilde{q}\cdot(p_b-p_{\bar b})\, \epsilon(P,\tilde{q},p_b,p_{\bar b}) \,\,
\xrightarrow[]{lab}\,\, \propto \,\, \vec{p}_{beam}\cdot (\vec{p}_b -\vec{p}_{\bar b})  \,\vec{p}_{beam}\cdot (\vec{p}_b\times\vec{p}_{\bar b}).
\end{eqnarray}

\item Multi-jet events, CP-odd correlations. Jets labelled without and with a ``prime'' are associated with the $b$ and $\bar{b}$ jets respectively. Notice that all one needs is to group each $b$ jet with two non-$b$ jets, but it is not necessary to actually distinguish the $b$ jet from the $\bar{b}$ jet.
\begin{eqnarray}
{\cal O}_{5jj} &=& \epsilon(p_b,p_{\bar b},p_{j1},p_{j1'})  \,\,
\xrightarrow[]{b\bar b ~CM}\,\, \propto \,\, \vec{p}_b \cdot (\vec{p}_{j1} \times \vec{p}_{j1'})
\nonumber \\
{\cal O}_{6jj} &=&  \epsilon(p_b,p_{\bar b},p_{j1}+p_{j2},p_{j1'}+p_{j2'})\,\,
\xrightarrow[]{t\bar t ~CM}\,\, \propto \,\, (\vec{p}_{j1}+\vec{p}_{j2})\cdot (\vec{p}_b \times \vec{p}_{\bar b}).
\end{eqnarray}

\end{itemize}

Additional examples can be found in Ref.~\cite{gupta} or can be easily constructed. Some observables have significantly higher sensitivity than others as can be inferred from the numerical results in these two references.

\section{Sources of CP violation}

We motivate the searches by presenting two examples of new physics with CP violation that generate some of the correlations discussed so far. We first consider anomalous top-quark couplings. This case will generate the most general T-odd correlations involving momenta from the production and decay processes. It will provide examples of both CP-odd and CP-even observables \cite{Antipin:2008zx}. The disadvantage of this case is that the asymmetries are small by assumption, because the anomalous couplings are necessarily small (to remain a valid description of the top-quark couplings).

As a second example we consider CP violation in extended scalar sectors. In this case the CP violation shows up as physical neutral scalars (color singlets or color octets) with simultaneous scalar and pseudo-scalar couplings to top-quark pairs and to gluons. This type of new physics can generate large intrinsic asymmetries (we have found some as large as 13\%) but they are hard to extract as they are concentrated in the range of $t\bar{t}$ invariant mass around the new resonance \cite{He:2011ws}. Also, the scalar nature of the new resonances prevents any correlations involving the beam momentum.

\subsection{Top-quark anomalous couplings}

\begin{figure}
\hspace{0.5in}\includegraphics[width=5in]{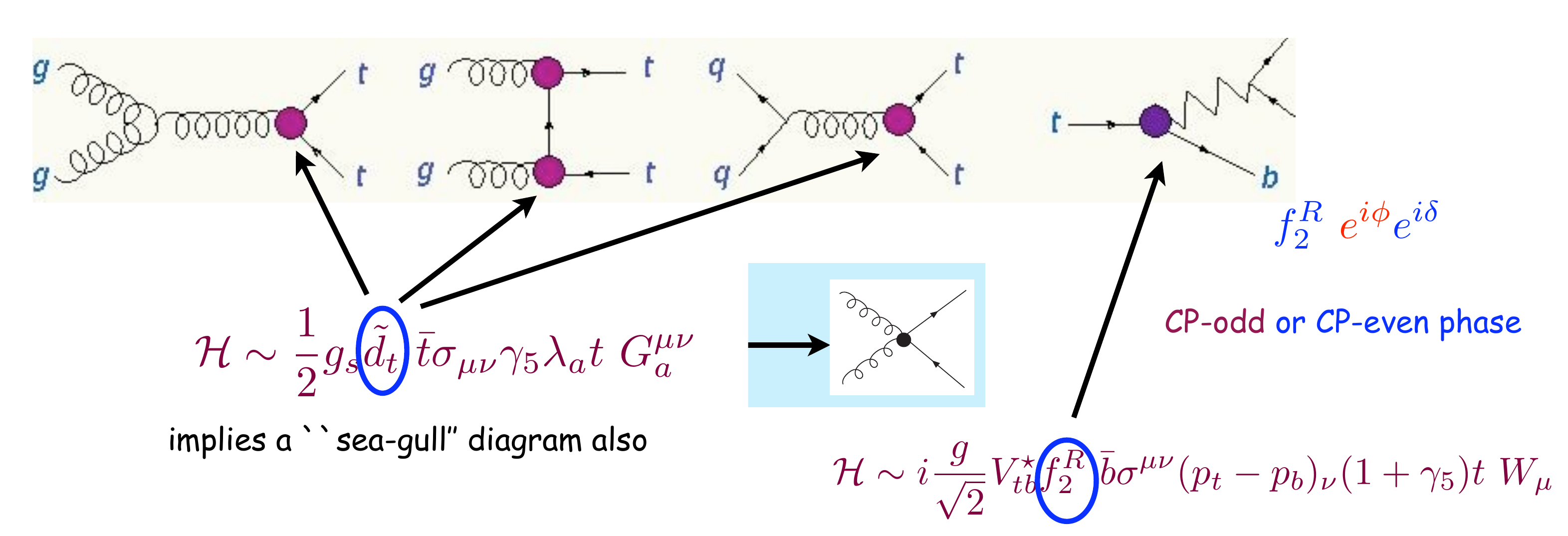}   
\caption{Anomalous couplings that induce CP-odd triple product correlations: $\tilde{d}_t$ (top CEDM)  and $f_2^R$.}
\label{fig2}
\end{figure}

In this case we modify the $ttg$ and $tbW$ interactions relevant for top-quark pair production and decay as in Figure~\ref{fig2}. For CP violation in the production vertex there is only one anomalous coupling, the top-quark color electric dipole moment (CEDM), $\tilde{d}_t$. For the effective $tbW$ interaction several anomalous couplings are possible but only one, $f_2^R$, can generate triple product correlations that are not suppressed by powers of the $b$-quark mass \cite{anomalous}.

The coupling $\tilde{d}_t$ is negligibly small in the SM but can be much larger in general. In particular, models in which CP violation is induced by the exchange of neutral scalars have contributions to the EDM (or CEDM) of fermions that scale as $m_f^3$, suggesting a potentially large top-quark CEDM. The CEDM of the top-quark is described by the effective Lagrangian as in Figure~\ref{fig2},
which modifies the $ttg$ vertex but also introduces a ``seagull'' $ttgg$ term that is required by gauge invariance. The contributions of these two vertices to top-quark production have been written down in a Lorentz invariant form \cite{Antipin:2008zx}. For example, for the leptonic decay of both $t$ and $\bar{t}$ this coupling produces a CP-odd part of the differential cross-section of the form
\begin{eqnarray}
d\sigma &\sim& C_1(s,t,u)\, \epsilon(p_t,p_{\bar{t}},p_{\mu^+},p_{\mu^-})  + C_2(s,t,u)\,(t-u) \,\epsilon(p_{\mu^+},p_{\mu^-},P,q) \nonumber \\
&+&C_3(s,t,u)\,(t-u) \,\left( P \cdot p_{\mu^+} \,
\epsilon(p_{\mu^-},p_t,p_{\bar{t}},q)+P \cdot p_{\mu^-} 
\,\epsilon(p_{\mu^+},p_t,p_{\bar{t}},q) \right)
\end{eqnarray}
where the sum and difference of parton momenta are denoted by $P$ and $q$ respectively. This formula can be easily adapted to semileptonic and purely hadronic channels. For example for $W$'s reconstructed as one jet the lepton momentum is replaced by the $b$-jet; for hadronic $W$ decay it is replaced by the $d$-jet momentum (which can be traded for the hardest non-$b$ jet), etc \cite{gupta}. Notice that the correlations in which $q$ appears are quadratic in $q$, as is necessary for two  indistinguishable initial state particles. The form factors $C_i$ can be found in Ref.~\cite{Antipin:2008zx} but are more conveniently evaluated numerically. This expression with three independent CP-odd and T-odd correlations appears to be the most general one, although a formal proof is not available. 

Additional $tbW$ anomalous couplings can introduce CP violation in the decay vertex, which we write as
\begin{eqnarray}
\Gamma^\mu_{Wtb} &=& 
-\frac{g}{\sqrt{2}} \, V_{tb}^\star \,\bar{u}(p_b) \left[ \gamma_\mu (f_1^L P_L+f_1^R P_R)-
i  \sigma^{\mu\nu} (p_t-p_b)_\nu (f_2^L P_L+f_2^R P_R)\right] u(p_t). \label{ftilde}
\end{eqnarray}
This vertex can be derived from a dimension five effective Lagrangian as in Ref.~\cite{delAguila:2002nf}, but unlike the case of the top cedm, the effective Lagrangian does not generate other vertices that affect this calculation. Numerically we use $V_{tb}\equiv 1$, $f_1^L=1$, $f_1^R=0$ and $f_2^L=0$ as in the SM, and allow for new physics only through the coupling $f_2^R$ which is the only one that can interfere with the SM to produce $T$-odd correlations that are not suppressed by the $b$-quark mass. To generate  $T$-odd observables the coupling $f_2^R$ must have a phase but this phase does not have to be $CP$ violating. We thus write $f_2^R=f\exp{i(\phi_f+\delta_f)}$ using $\phi_f$ to parametrize a $CP$ violating phase due to new physics and $\delta_f$ a $CP$ conserving phase arising from real intermediate states at the loop level. The coupling $f_2^R$ occurs already in one-loop QCD corrections to the $tbW$ vertex with a value of 0.03 \cite{qcdcorr} but without a phase. A unitarity phase occurs in processes with additional gluons \cite{Hagiwara:2007sz}.

The spin and color averaged matrix element squared containing the $T$-odd correlations in this case looks like~\cite{Antipin:2008zx} \begin{eqnarray}
|{\cal M}|^2_{T} &=& \, 
f\sin(\phi_f+\delta_f)\, \epsilon(p_t,p_{ b},p_{\ell^+},Q_{t}) +
 f\sin(\phi_f-\delta_f) \,\epsilon(p_{\bar t},p_{ \bar{b}},p_{\ell^-},Q_{\bar{t}}) .
 \label{asymcpdec}
\end{eqnarray} 
All the terms in Eq.~\ref{asymcpdec} contain three four-momenta from one of the decay vertices and a fourth ($Q$) `spin-analyzer' which is a linear combination of other momenta in the reaction (its precise form can be found in Ref.~\cite{Antipin:2008zx}). Note that these correlations are not CP-odd as was the case for CP violation in the production vertex. One needs to compare the top and anti-top decays to extract either a CP-odd observable or a CP-even one. The analytic form, Eq.~\ref{asymcpdec} guides our expectations for the asymmetries but it is again easier to perform all calculations numerically as described next.

\section{Numerics}

We performed a basic simulation to determine the statistical sensitivity of LHC to the anomalous couplings using MadGraph \cite{madgraph} to generate all signal and background events. Our basic result is that the LHC with 10~fb$^{-1}$ at 14~TeV has a potential $5\sigma$ statistical sensitivity to integrated asymmetries $A_i$ of $3\%$. This sensitivity can be translated into bounds on the anomalous couplings as listed in Table~\ref{tab:limits}. At 7 or 8 TeV where the $t\bar{t}$ cross-section is about five times smaller, the statistical sensitivity should be roughly 2.5 times worse for the same 10~fb$^{-1}$.

An important feature is that there are no background issues for these CP studies beyond those already present in the selection of top-quark pair events. This is because all known backgrounds  are CP conserving. Residual background after event selection will dilute the statistical sensitivity of the signals but will not fake them. It is important to carry out further detector level simulations  to identify potential sources of systematic error. Details of our numerical simulations can be found in the original papers, here we summarize the best results for LHC and put them in perspective in Table~\ref{tab:limits}. 

\begin{table}
  \caption{Sensitivity limits at LHC (14 TeV) compared to sample models.
                           }
  \vspace{0.1 in}
  \label{tab:limits}
  \begin{tabular}{|l|l|l|}
  
    \hline
    	 coupling & $\tilde{d}\left[\frac{1}{m_t}\right]$ & $\tilde{f}\left[\frac{1}{m_t}\right]$ \\
	 \hline
      Theory estimate & $< 10^{-13}$ SM \cite{review} & 0.03 QCD  \cite{qcdcorr}   \\
      &  $\sim 10^{-3}$ SUSY \cite{review} & CP conserving, no phases. \\
      & $\sim 0.003$ vector-like multiplets \cite{Ibrahim:2011im} & Phases require extra gluons \cite{Hagiwara:2007sz} \\
        \hline
$5\sigma$ sensitivity with 10 fb$^{-1}$ & 0.05     &0.10 \\
\hline
  \end{tabular}
\end{table}

The QCD estimate is for the magnitude of $f$, without any phases. At this level the coupling cannot produce T-odd correlations but if we assume that there are large unitarity phases this number is a rough estimate for the level of the CP even T-odd correlations that appear in the SM. Absorptive phases arise at one-loop in QCD in processes with an additional gluon, and this has been considered in  detail in \cite{Hagiwara:2007sz}. 

\begin{figure}
\vspace{0.2in}
\hspace{1.0in}\includegraphics[width=4.0in]{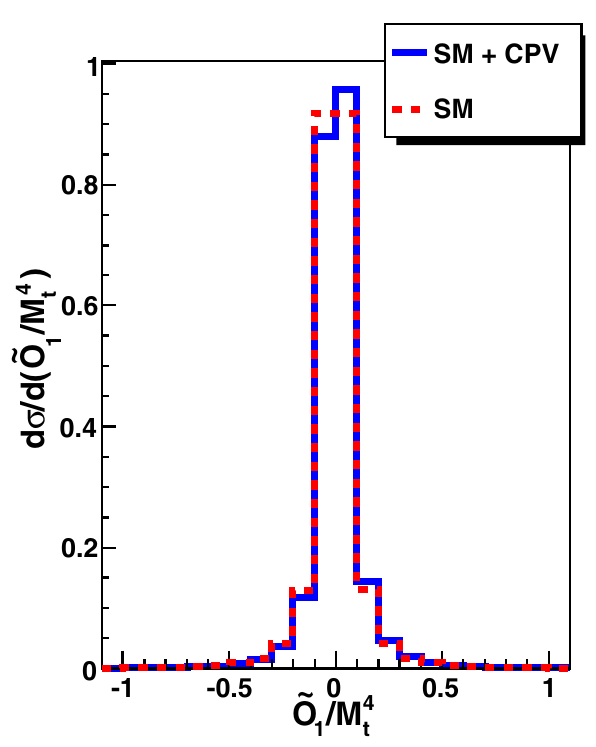}     
\caption{Differential cross-section as a function of the T-odd correlation ${\cal O}_1$. Terms linear in this correlation generate an asymmetry that signals CP violation.}
\label{f:asym}
\end{figure}

As an example of the asymmetries in differential distributions induced by the CP violating terms we show in Figure~\ref{f:asym} the effect of $\tilde{d}=5\times10^{-4}$~GeV~$^{-1}$ on $d\sigma/d{\cal O}_1$.

\subsection{Extended scalar sectors}

\begin{figure}
\vspace{0.2in}
\includegraphics[width=6.0in]{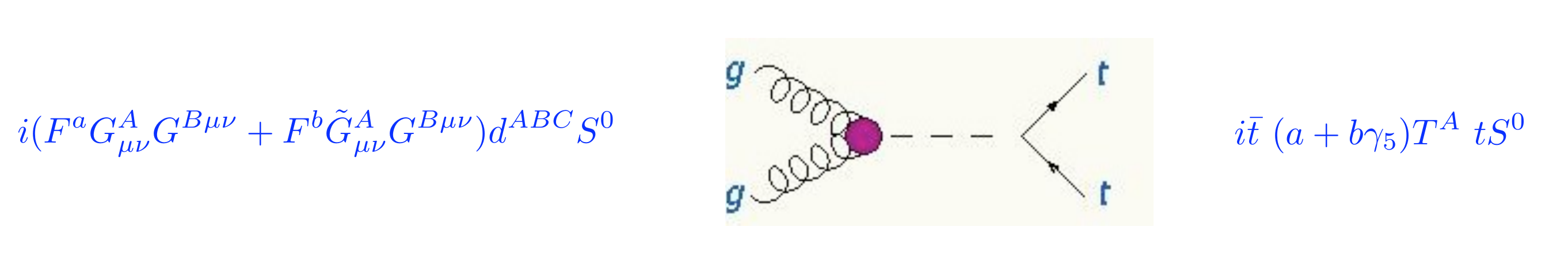}     
\caption{Effective scalar/pseudo-scalar couplings of new resonance.}
\label{f:news}
\end{figure}

In order to show that it is possible to have large CP violating asymmetries within specific models we now consider extended scalar sectors in which there is a color octet, electroweak doublet \cite{coloroctet}. Related studies for color singlets also abound in the literature \cite{extrahiggs}. This specific multiplet is motivated by minimal flavor violation and has been studied in some detail recently. There are two neutral (color octet) scalars $S_R^0$ and $S_I^0$ which in the general case with CP violation have both scalar and pseudo-scalar couplings to the top-quark as well as to gluons. The model admits CP violation through two new couplings: a phase $\eta_U$ in the coupling to top-quarks of the form \cite{coloroctet}
\begin{eqnarray}
{\cal L} = -\frac{\sqrt{2}}{v}\eta_U\ e^{i\alpha_U}\ \bar U_RT^A\hat{M}^uU_L\ S^{A0} + {\rm ~h.c.}+\cdots
\end{eqnarray}
and a phase $\alpha_4$ in the scalar potential \cite{coloroctet}
\begin{eqnarray}
V &=& {\lambda\over 4} \left ( H^{\dagger i} H_i - {v^2\over 2}\right )^2 + 2 m^2_s {\rm Tr}\ S^{\dagger i}S_i + \lambda_1 H^{\dagger i} H_i {\rm Tr}\ S^{\dagger j}S_j + \lambda_2 H^{\dagger i}H_j
{\rm Tr}\  S^{\dagger j}S_i \nonumber\\
&+& \left[ \lambda_3 e^{i \alpha_3} H^{\dagger i} H^{\dagger j} {\rm Tr}\ S_i S_j + \lambda_4 e^{i \alpha_4}H^{\dagger i}{\rm Tr}\ S^{\dagger j}S_j S_i +  \lambda_5e^{i \alpha_5} H^{\dagger i}{\rm Tr}\ S^{\dagger j}S_i S_j + h.c\right] \nonumber \\
&+&\lambda_6 {\rm Tr}\  S^{\dagger i} S_i S^{\dagger j}S_j +\cdots
\end{eqnarray}
The scalar potential appears to have room for three new phases, but one of them ($\alpha_3$ in our case) can be set to zero by convention and the other two are related by custodial symmetry $\alpha_4=-\alpha_5$. When these phases are non-zero, the couplings of both $S_{R,I}^0$ to $t\bar{t}$ and to $gg$ contain scalar and pseudo-scalar components as shown schematically in Figure~\ref{f:news}. This is what signals CP violation, and in this case generates only one of the T-odd correlations that appear in the general case, namely
\begin{equation}
d\sigma_{S^0_I,S^0_R} \sim C_{1,S^0_I,S^0_R}(s,t,u)\ {\cal O}_1.
\end{equation}
An explicit expression for $C_{1,S^0_I,S^0_R}(s,t,u)$ can be found in Ref.~\cite{He:2011ws}, but is unnecessary for numerical evaluation.

As before, we use MadGraph \cite{madgraph} to generate all signal and background events for the process $pp \to t\bar{t} \to b \bar{b} \mu^+\mu^-\nu\bar\nu$ at a 14~TeV LHC. We concentrate on the dimuon channel for simplicity but all channels can be studied with the corresponding observables discussed above. There are several salient features of this study \cite{He:2011ws}
\begin{itemize}
\item The intrinsic asymmetries can be rather large. An integrated asymmetry based on ${\cal O}_1$, for example, can reach a peak value of about $13\%$ at each resonance.

\item The asymmetries at each of the two resonances, $S_{R,I}^0$, have opposite signs, so that a completely integrated asymmetry vanishes.

\item The raw asymmetry is diluted by QCD produced top-quark pairs and can be rather small if the resonance does not stand out.

\end{itemize}

\begin{figure}
\vspace{0.2in}
\includegraphics[width=6.0in]{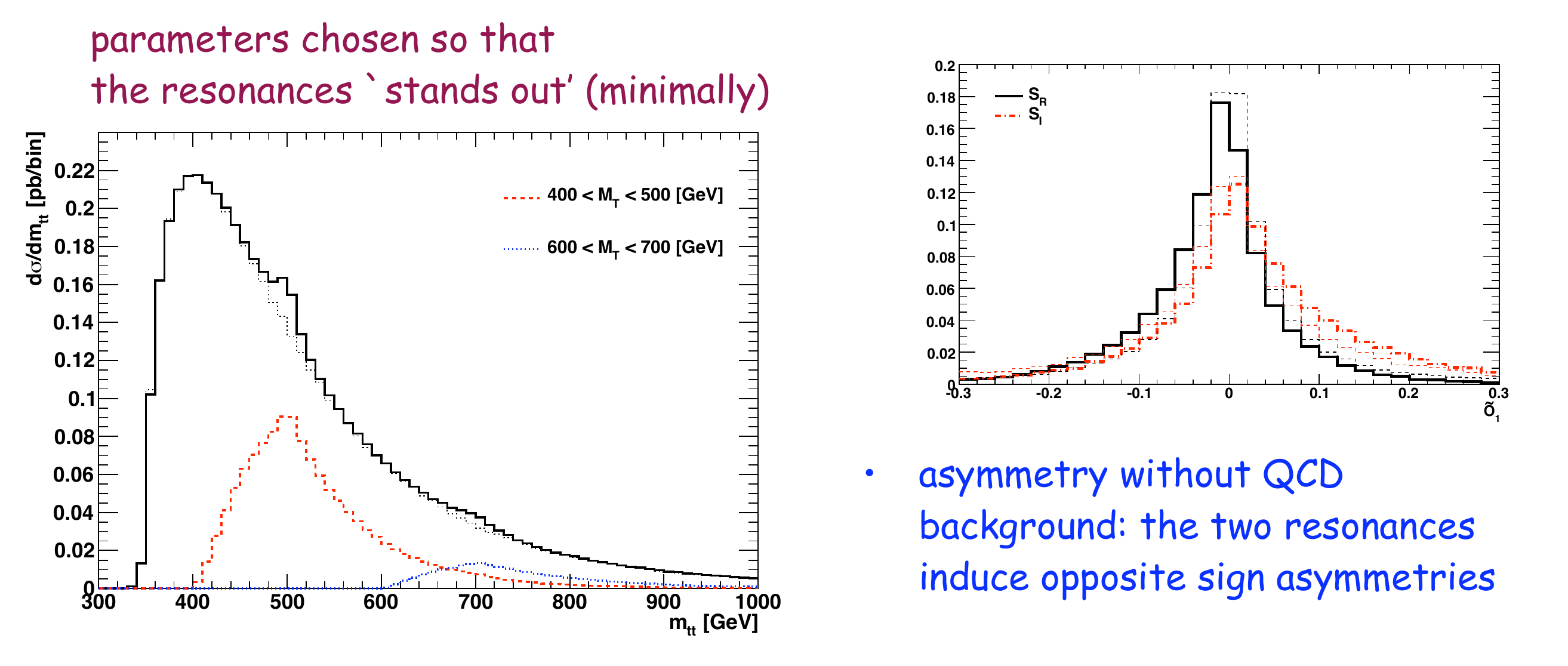}     
\caption{$d\sigma/dm_{t \bar{t}}$ distributions without and with contribution
 of scalars in Case~1$^\prime$, plotted in dotted and solid line,
 respectively, in an unit of pb per 5 GeV bin. Dashed lines are for the distribution after $400<M_{T}<500$~[GeV] (red
 long-dashed) and $600<M_{T}<700$~[GeV] (blue short-dashed) cuts.}
\label{color}
\end{figure}

The model has several free parameters, but if we choose them so that they produce resonances at 500 and 700 GeV that stand out minimally above the QCD continuum, the resulting invariant mass distribution looks as Figure~\ref{color} \cite{He:2011ws},
The parameters used are labelled Case~1$^\prime$ after Ref.~\cite{He:2011ws}, for which 
\begin{eqnarray}
m_R=500,\ , m_I=700~{\rm GeV},\, 
 \eta_U=3, \, \lambda_{4,5}=1,\, \alpha_u=\pi/4\, 
\end{eqnarray}
and these result in resonant decay widths $S_R=24.3$, $S_I=47.7$ GeV; production cross-sections $S_R=60.4$, $S_I=24$ fb and raw asymmetries $A_1$ around each resonance of $S_R\to -0.127$, $S_I\to 0.103$ respectively. Using the $M_T$ cuts in Figure~\ref{color} to separate the resonances, results in asymmetries that are significantly diluted. To keep the dilution minimal one must be able to cleanly cut the (narrow) regions around the resonances. One can conclude from this that it would be necessary to first observe the resonance in order to search for this type of CP violation, but one can also draw the lesson that in a generic search it would be useful to study many asymmetries that are integrated over narrow bins of $m_{t\bar{t}}$ rather than the fully integrated counting asymmetry. How to divide these bins would require a balance between increasing statistics for sensitivity versus narrowing the bin to pick up new physics. Motivated by the LHC, several additional  studies have appeared recently\cite{recent} discussing different observables.

\section{Conclusions}

The LHC is entering an era of precision measurements in the top-quark system and we have argued that the search for CP violation should be an integral part of this program. Any observation would signal new physics.

We have studied several T-odd correlations that illustrate the different possibilities in searching for CP violation in top physics corresponding to the different decay channels and to the option of separating unitarity phases.

We have motivated these searches with two examples of new physics that would lead to observable CP violating signals. Anomalous top quark couplings illustrate the general case and extended scalar sectors illustrate potentially large intrinsic asymmetries.

There exists a recent D0 thesis studying these asymmetries that finds no systematic problems within the limited statistics available to them \cite{thesis}. We urge the ATLAS and CMS collaborations to carry out these studies at the LHC.

\acknowledgments
This talk is based on work done with  John~F.~Donoghue, Oleg~Antipin, Sudhir Gupta, Sehwook~Lee, Xiao-Gang~He and Hiroshi Yokoya. I also wish to thank the organizers of HQL2012 for a very stimulating meeting.

\end{document}